\documentstyle[12pt]{article}
\begin {document}
\baselineskip 2.25pc

\begin{center}
{\Large Levinson theorem for Dirac particles in two dimensions}
\vskip 2pc
 Qiong-gui Lin\\
        China Center of Advanced Science and Technology (World
	Laboratory),\\
        P.O.Box 8730, Beijing 100080, People's Republic of China\\
        and\\
        Department of Physics, Zhongshan University, Guangzhou
        510275,\\
        People's  Republic of China \footnote{Mailing address}

\end{center}
\vfill
\centerline{\bf Abstract}

The  Levinson theorem for nonrelativistic
quantum mechanics in two spatial dimensions is generalized to Dirac
particles moving in a central field. The theorem relates the total
number of bound states with angular momentum $j$ ($j=\pm 1/2, \pm 3/2,
\ldots $), $n_j$, to the phase shifts $\eta_j(\pm E_k)$ of scattering
states at zero momentum as follows: $\eta_j(\mu)+\eta_j(-\mu)=
n_j\pi$.

\vfill

\leftline {PACS number(s): 34.10.+x, 03.65.-w, 11.80.-m}

\newpage

\section*{\centerline{\large I. Introduction}}

In 1949, Levinson established a theorem in nonrelativistic quantum
mechanics[1]. The theorem gives a
relation between bound states and scattering states in a given
angular momentum channel $l$, i.e.,  the total number of bound states
$n_l$ is related to the phase shift $\delta_l(k)$ at threshold
($k=0$):
$$\delta_l(0)=n_l\pi,\quad l=0,1,2,\ldots.$$
The case $l=0$ should be modified as
$$\delta_0(0)=(n_0+1/2)\pi$$
when there exists a zero-energy resonance (a half bound state)[2].
This is one of the most interesting
and beautiful results in nonrelativistic quantum theory. The subject
has been studied by many authors (some are listed in the Refs.[2-8])
and  generalized to relativistic quantum mechanics[6,9-14].
However, most of these authors deal with
the problem in ordinary three-dimensional space. 
A two-dimensional version of Levinson's theorem
does not appear to have been discussed by previous
authors. In view of the wide interest in lower-dimensional field
theories in recent years, e.g., Chern-Simons theory in 2+1
dimensions, and the previous applications of Levinson's theorem
to field theories[15], it seems of interest to study the theorem
in two-dimensional space. On the other hand, as the problem exhibits
some new features in two spatial dimensions, it may also be of
interest in its own right.
We are thus led to consider the problem.

In a recent work[16] we have established the Levinson theorem in
two spatial dimensions, which takes the following form:
$$\eta_m(0)=n_m\pi, \quad m=0,1,2,\ldots$$
where  $\eta_m(0)$ is the phase shift of the $m$th partial wave
at threshold, and $n_m$ is  the total
number of bound states with angular momentum $m$ (it also equals the
total number of bound states with angular momentum $-m$ when $m\ne0$). 
As in three dimensions, the modulo-$\pi$ ambiguity in the definition
of $\eta_m(k)$ has been resolved by setting $\eta_m(\infty)=0$ (rather
than a multiple of $\pi$) which can be freely done in nonrelativistic
theory.
The theorem is similar to the three-dimensional one but simpler in
that the existence of half bound states (possible for $m=0,1$) does
not alter the form of the theorem.

In this paper we extend the previous work to the relativistic
case and establish the Levinson theorem for Dirac particles moving
in an external central field in two spatial dimensions. In a given
angular momentum channel $j$ ($j=\pm1/2, \pm3/2,\ldots$), the theorem
relates the total number of bound states $n_j$ to the phase shifts
$\eta_j(\pm E_k)$ at zero momentum:
\begin{equation}
\eta_j(\mu)+\eta_j(-\mu)=n_j\pi.
\end{equation}
In the relativistic theory, one is not allowed to set $\eta_j(\pm
\infty)=0$. But the modulo-$\pi$ ambiguity in the definition of
$\eta_j(\pm E_k)$ may be appropriately resolved (see Sec. V). As in
the nonrelativistic case, the theorem is similar to the
three-dimensional one[10] but somewhat simpler. In three dimensions,
the theorem should be modified when there exists a half bound state,
but here we have no such trouble.

Throughout this paper natural units where $\hbar=c=1$ are employed.
In the next section we first discuss various aspects of the solutions
of the Dirac equation in an external central field in two spatial
dimensions. Then we give a brief formulation of the partial-wave
method for potential scattering of Dirac particles.
In Sec. III the behavior of the phase shifts near $k=0$ is analyzed.
In Sec. IV we establish the Levinson theorem using the Green function
method[3,4,6,10]. Sec. V is devoted to some discussions 
relevant to the theorem.

\section*{\centerline{\large II. Dirac particles in an external central
field in two dimensions}}

We work in (2+1)-dimensional space-time. The Dirac equation in an
external vector field $A_\nu(t,{\bf r})$ reads
\begin{equation}
(i\gamma^\nu D_\nu-\mu)\Psi=0,
\end{equation}
where $\mu$ is the mass of the particle, $D_\nu=\partial_\nu+ieA_\nu$,
$e$ is the coupling constant, and summation over the repeated Greek
index $\nu$ ($\nu=0,1,2$) is implied. The $\gamma^\mu$ are Dirac
matrices satisfying the Clifford algebra:
\begin{equation}
\{ \gamma^\mu, \gamma^\nu \}=2g^{\mu\nu},
\end{equation}
where $g^{\mu\nu}={\rm diag}(1,-1,-1)$ is the Minkowskian metric.
In this paper we only consider the zero component of $A_\nu$,
which is cylindrically symmetric, namely, we consider the special
case where
\begin{equation}
{\bf A}=0, \quad eA_0=V(r),
\end{equation}              
where $r$ is one of the polar coordinates $(r,\theta)$ 
in two-dimensional space. In this case we may set
\begin{equation}
\Psi(t, {\bf r})=e^{-iEt}\psi({\bf r}),
\end{equation}
and get a stationary equation for $\psi({\bf r})$:
\begin{equation}
H\psi=E\psi,
\end{equation}
where the Hamiltonian
\begin{equation}
H={\bf\alpha\cdot p}+\gamma^0\mu+V(r),
\end{equation}           
where ${\bf p}=-i\nabla$, ${\bf \alpha}=\gamma^0{\bf \gamma}$
or $\alpha^i=\gamma^0\gamma^i$ ($i=1,2$).

In two dimensional space, the orbital angular momentum has only one
component:
\begin{equation}
L=\epsilon^{ij}x^ip^j,
\end{equation}
where $\epsilon^{ij}$ is antisymmetric in $i$, $j$ and $\epsilon^
{12}=1$, and summation over the repeated Latin indices $i$, $j$
is implied. It is easy to show that $[L,H]=i\epsilon^{ij}\alpha^i
p^j\ne0$, thus $L$ is not a constant of motion even when $V=0$. Let
\begin{equation}
S={i\over 4}\epsilon^{ij}\gamma^i\gamma^j.
\end{equation}
It is easy to show that $[S,H]=-i\epsilon^{ij}\alpha^ip^j$, then
$[L+S,H]=0$, and the quantity
\begin{equation}
J=L+S
\end{equation}      
is a constant of motion. It is natural to regard $J$ as the total
angular momentum and $S$ as the spin angular momentum in two
dimensions.

\section*{\centerline{\large II.A. Solutions in the external
central field}}

To solve Eq.(6) a representation of the Dirac matrices is necessary.
This can be realized by the Pauli matrices:
\begin{equation}
\gamma^0=\sigma^3,\quad \gamma^1=i\sigma^1,\quad \gamma^2=i\sigma^2.
\end{equation}
In this representation $S=\sigma^3/2$. Let
\begin{equation}
\psi_j(r,\theta)=\left(
\begin{array}{c}
F(r)e^{i(j-1/2)\theta}/\sqrt{2\pi}\\    \\
G(r)e^{i(j+1/2)\theta}/\sqrt{2\pi}\end{array}
\right)=\left(
\begin{array}{c}
v_1(r)e^{i(j-1/2)\theta}/\sqrt{2\pi r}\\  \\
v_2(r)e^{i(j+1/2)\theta}/\sqrt{2\pi r}\end{array}
\right),\quad j=\pm1/2,\pm3/2,\ldots.
\end{equation}      
It is easy to show that $\psi_j$ is an eigenfunction of $J$:
\begin{equation}
J\psi_j=j\psi_j.
\end{equation}         
So that $\psi_j$ has angular momentum $j$.
(There is no danger of confusing the angular momentum with the Latin
index $j$ used above, as the latter will not appear henceforth.)
The radial wave functions $F$ and $G$ satisfy the system of equations
$$F'-{j_-\over r}F+(E+\mu-V)G=0,\eqno(14{\rm a})$$
$$G'+{j_+\over r}G-(E-\mu-V)F=0,\eqno(14{\rm b})$$
where $j_\pm=j\pm1/2$, and primes denote differentiation with respect
to argument. We will have occasions to use the system of equations
for $v_1$ and $v_2$, so we also write down it here:
$$v_1'-{j\over r}v_1+(E+\mu-V)v_2=0,\eqno(15{\rm a})$$
$$v_2'+{j\over r}v_2-(E-\mu-V)v_1=0.\eqno(15{\rm b})$$
\addtocounter{equation}{2}
From Eq.(14a), we have
\begin{equation}
G=-{F'-(j_-/r)F\over E+\mu-V}.
\end{equation}
Substituting this into Eq.(14b) we get an equation for $F$ alone:
\begin{equation}
F''+\left({1\over r}+{V'\over E+\mu-V}\right)F'+\left[E^2-\mu^2
-{j_-^2\over r^2}-2EV+V^2-{j_-V'\over r(E+\mu-V)}\right]F=0.
\end{equation}       
With a given potential $V(r)$ that is regular everywhere except
possibly at $r=0$
and appropriate boundary conditions
one can in principle solve Eq.(17) for F, and get $G$ from
Eq.(16). Equivalent to Eqs.(16) and (17), we may have
\begin{equation}
F={G'+(j_+/r)G\over E-\mu-V},
\end{equation}
\begin{equation}
G''+\left({1\over r}+{V'\over E-\mu-V}\right)G'+\left[E^2-\mu^2
-{j_+^2\over r^2}-2EV+V^2+{j_+V'\over r(E-\mu-V)}\right]G=0.
\end{equation}
If in some region or at some point $E+\mu-V=0$,
we can not use Eqs.(16) and (17).
But then $E-\mu-V\ne0$ in that region or in the neighbourhood of that
point, and we can use Eqs.(18) and (19). Indeed, the fundamental
equation is Eq. (14) or Eq. (15). They are regular everywhere except
at $r=0$. In principle one can solve them by direct integration
without the help of Eqs. (16-19). Eqs. (16-17) or Eqs. (18-19) are
to be employed when convenient. The possible singularities in these
equations due to the vanishing denominator $E+\mu-V$ or
$E-\mu-V$ should not cause any trouble in principle. Nevertheless,
attention should be paid to these possible singularities when we
have to use Eq. (17) or Eq. (19)  in the whole range of $r$.

For free particles, $V=0$, then Eq.(17) becomes
\begin{equation}
F''+{1\over r}F'+\left(E^2-\mu^2-{j_-^2\over r^2}\right)F=0.
\end{equation}                           
In order to get well behaved solutions one should have $E^2-\mu^2
\ge0$. Thus we have positive-energy solutions with $E\ge\mu$ and
negative-energy solutions with $E\le-\mu$. Let us define
$k=\sqrt{E^2-\mu^2}\ge0$, and denote positive-(negative-)energy
solutions by the subscript $k$ ($-k$), thus we have, say,
\begin{equation}
E_{\pm k}=\pm E_k=\pm \sqrt{k^2+\mu^2}.
\end{equation}
It is not difficult to find the following solutions for free
particles:
$$F_{\pm kj}^{(0)}(r)=\sqrt{(E_k\pm \mu)k\over 2E_k}
J_m(kr),\eqno(22{\rm a})$$
$$G_{\pm kj}^{(0)}(r)=\pm \epsilon(j)\sqrt
{(E_k\mp\mu)k\over 2E_k}
J_{m+\epsilon(j)}(kr), \eqno(22{\rm b})$$
\addtocounter{equation}{1}
where $m=|j_-|$, $J_m(kr)$ is the Bessel function,  $\epsilon(j)=1
(-1)$ when $j>0(<0)$, and the normalization factors are chosen such
that the orthonormal relation takes the form
\begin{equation}
\int d{\bf r}\,\psi_{\pm k'j'}^{(0)\dagger}({\bf r})
\psi_{\pm kj}^{(0)}({\bf r})=\delta(k-k')\delta_{jj'},
\end{equation}
$$
\int d{\bf r}\,\psi_{\mp k'j'}^{(0)\dagger}({\bf r})
\psi_{\pm kj}^{(0)}({\bf r})=0. \eqno(23')
$$
The completeness of these solutions is ensured by the following
relation which can be verified straightforwardly:
\begin{equation}
\sum_j\int_0^\infty dk\, [\psi_{kj}^{(0)}({\bf r})\psi_{kj}^{(0)
\dagger}({\bf r'})+
\psi_{-kj}^{(0)}({\bf r})\psi_{-kj}^{(0)
\dagger}({\bf r'})]=\delta ({\bf r}-{\bf r'}).
\end{equation}
When $r\to \infty$, the radial wave functions have the asymptotic
form:
$$v_{\pm kj1}^{(0)}(r)\to \sqrt{E_k\pm \mu\over \pi E_k}
\cos \left(kr-{m\pi\over2}-{\pi\over4}\right),
\eqno(25{\rm a})$$
$$v_{\pm kj2}^{(0)}(r)\to \pm\sqrt{E_k\mp \mu
\over \pi E_k}
\sin \left(kr-{m\pi\over2}-{\pi\over4}\right).
\eqno(25{\rm b})$$
\addtocounter{equation}{1}
This can be obtained by using Eq.(22) and the asymptotic formula
for the Bessel function:
\begin{equation}
J_n(x) \stackrel{x\to\infty}{\longrightarrow}
\sqrt{2\over\pi x}\cos \left(x-{n\pi\over2}-{\pi\over4}\right).
\end{equation}   

Now we consider particles moving in the external central potential
$V(r)$. We assume that $V(r)\to 0$ more rapidly than $r^{-2}$ when
$r\to \infty$, and is less singular than $r^{-1}$ when $r\to 0$.
Then, for very large $r$, Eq.(17) takes the same form as Eq.(20).
It is easy to see that $E^2\ge\mu^2$ gives scattering solutions
while $E^2<\mu^2$ gives bound state solutions (bound states with
$E=\pm \mu$ are also possible, see Sec.V). Scattering states will be
denoted as above, while bound states will be denoted by a
subscript $\kappa$ which takes discrete values. The orthonormal
relations are given by
\begin{equation}
\int d{\bf r}\,\psi_{\pm k'j'}^\dagger({\bf r})\psi_{\pm kj}
({\bf r})=\delta(k-k')\delta_{jj'},
\end{equation}  
$$
\int d{\bf r}\,\psi_{\kappa'j'}^\dagger({\bf r})\psi_{\kappa j}
({\bf r})=\delta_{\kappa \kappa'}\delta_{jj'},
\eqno(27')
$$
and vanishing ones similar to Eq.$(23')$. The completeness
relation is similar to Eq.(24) but has an additional term
on the left-hand side (lhs):
\begin{equation}
\sum_j\int_0^\infty dk\, [\psi_{kj}({\bf r})\psi_{kj}^
\dagger ({\bf r'})+
\psi_{-kj}({\bf r})\psi_{-kj}^
\dagger({\bf r'})]+\sum_{\kappa j}\psi_{\kappa j}({\bf r})
\psi_{\kappa j}^\dagger({\bf r'})
=\delta ({\bf r}-{\bf r'}).
\end{equation}    

As pointed out above, Eq.(17) takes the form of Eq.(20) at large
$r$, so the solution $F_{kj}(r)$ is giver by a linear combination
of $J_m(kr)$ and $N_m(kr)$, the Neumann function, at large $r$.
Using Eq.(26) and
\begin{equation}
N_n(x) \stackrel{x\to\infty}{\longrightarrow}
\sqrt{2\over\pi x}\sin \left(x-{n\pi\over2}-{\pi\over4}\right),
\end{equation}  
and Eq.(16), the asymptotic forms for the radial wave functions
can be shown to be
$$v_{\pm kj1}(r)\to \sqrt{E_k\pm \mu\over \pi E_k}
\cos \left[kr-{m\pi\over2}-{\pi\over4}+
\eta_j(\pm E_k)\right],\eqno(30{\rm a})$$
$$v_{\pm kj2}(r)\to \pm\sqrt{E_k\mp \mu
\over \pi E_k}
\sin \left[kr-{m\pi\over2}-{\pi\over4}+
\eta_j(\pm E_k)\right], \eqno(30{\rm b})$$
\addtocounter{equation}{1}
when $r\to\infty$, where $m=|j_-|$ as before, $\eta_j(\pm E_k)$
are the phase shifts. They depend on $j$ rather than $|j|$, and
also depend on the sign (not only the magnitude) of the energy,
as Eq.(17) does. Compared with Eq.(25), the asymptotic
forms in the external certral field are distorted by the phase
shifts. But it should be remarked that the normalization factors
in Eq.(30) are the same as in Eq.(25).

\section*{\centerline{\large II.B. Partial-wave analysis of
scattering by the central field}}

It is well known that positive-(negative-)energy solutions
correspond to particles (antiparticles) after second quantization.
In this subsection we discuss the scattering of positive-energy
solutions by the central field $V(r)$ described above. The
scattering of negative-energy solutions can be formally discussed
in a similar way.

The probability current density associated wiht Eq.(2) is given by
\begin{equation}
{\bf j}=\psi^\dagger{\bf\alpha}\psi.
\end{equation}
The incident wave may be chosen as
\begin{equation}
\psi_{in}=\left(
\begin{array}  {c}
i\sqrt{\displaystyle{E_k+\mu\over 2E_k}} \\ \\
\sqrt{\displaystyle{E_k-\mu\over 2E_k}}
\end{array} 
\right)e^{ikx}, \quad k>0,
\end{equation} 
which is a solution of Eq.(6) with positive energy $E_k$ when
$x\to -\infty$. The incident probability current density is
\begin{equation}
{\bf j}_{in}={\bf e}_xk/E_k={\bf e}_xv,
\end{equation}             
where ${\bf e}_x$ is the unit vector in the $x$ direction, the
incident direction in this case. The scattered wave should have the
asymptotic form when $r\to \infty$:
\begin{equation}
\psi_{sc}\to \sqrt{i\over r}\left(
\begin{array}  {c}
f_1(\theta) \\ \\  f_2(\theta)
\end{array} 
\right)e^{ikr},
\end{equation}   
where the factor $\sqrt i=e^{i\pi/4}$ is introduced for latter
convenience.  Then the $r$ and $\theta$ components of ${\bf j}_
{sc}$ at large $r$ can be shown to be
\begin{equation}
j_{sc\, r}={2\over r}{\rm Im}(e^{i\theta}f_1f_2^*), \quad
j_{sc\, \theta}={2\over r}{\rm Re}(e^{i\theta}f_1f_2^*).
\end{equation}
However, the expression (34) does not solve the Dirac equation at
large $r$ if $f_1(\theta)$ and $f_2(\theta)$ are independent of
each other.  In order to satisfy Eq.(6) at large $r$, one should
have
\begin{equation}
f_2(\theta)=-{ik\over E_k+\mu}e^{i\theta}f_1(\theta).
\end{equation}   
Setting
\begin{equation}
f_1(\theta)=i\sqrt{E_k+\mu\over 2E_k}f(\theta),
\end{equation}
it is easy to show that
\begin{equation}
{\bf j}_{sc}={\bf e}_rv|f(\theta)|^2/r
\end{equation}
when $r\to\infty$, where ${\bf e}_r$ is the unit vector in the radial
direction. Thus the differential cross section (in two dimensions
the cross section may be more appropriately called cross width) is
given by
\begin{equation}
\sigma(\theta)=|f(\theta)|^2.
\end{equation}
The outgoing wave comprises Eqs.(32) and (34). With the relation
(36) and the definition (37), it takes the following asymptotic
form when $r\to\infty$:
\begin{equation}
\psi\to \left(
\begin{array}  {c}
i\displaystyle{\sqrt{E_k+\mu\over2E_k}\left[e^{ikx}+\sqrt
{\frac ir}e^{ikr}f(\theta)\right]}  \\ \\
\displaystyle{\sqrt{E_k-\mu\over2E_k}\left[e^{ikx}+\sqrt
{\frac ir}e^{ikr}e^{i\theta}f(\theta)\right]}
\end{array}  \right).
\end{equation}  
On the other hand, the solution of Eq.(6) with definite energy
$E_k>0$ has the form
\begin{equation}
\psi (r,\theta)=\sum_j a_j\psi_{kj}(r,\theta),
\end{equation}
where the summation is made over all $j=\pm1/2,\pm3/2,\ldots$.
The asymptotic form of Eq.(41) can be obtained from Eq.(30).
By using the formula
\begin{equation}
e^{ikx}=\sum_{n=-\infty}^{+\infty}i^{|n|}J_{|n|}(kr)
e^{in\theta}
\end{equation}                                      
and Eq.(26), one can compare Eq.(40) with the asymptotic form of
Eq.(41). They must coincide with each other for appropriately chosen
$a_j$'s. In this way one finds all $a_j$ in terms of $\eta_j(E_k)$
and
\begin{equation}
f(\theta)=\sum_j\sqrt{2\over\pi k}e^{i\eta_j(E_k)}
\sin\eta_j(E_k)e^{ij_-\theta}.
\end{equation}                                      
The total cross section $\sigma_t$ turns out to be
\begin{equation}
\sigma_t=\int_0^{2\pi}d\theta\,\sigma(\theta)={4\over k}
\sum_j\sin^2\eta_j(E_k).
\end{equation}
One easily realizes that all information of the scattering process
is contained in the phase shifts. The latter are determined by
solving the system of equations (15) with the boundary conditions
(30). The purpose of the Levinson theorem is to establish a
relation between scattering states and bound states, specifically,
to establish a relation between the phase shifts and the total
number of bound states in a given angular momentum channel $j$.

\section*{\centerline{\large III. Phase shifts near threshold}}

In this section we discuss the behavior of the phase shifts
$\eta_j(\pm E_k)$ near $k=0$. This will be employed in the
next section. For exact analysis let us cut off the potential.
That is, we consider potentials that satisfy $V(r)=0$ when $r>a>0$.
Such potentials will be denoted by $V_a(r)$ in the following. In
the region $r>a$, then, Eq.(17) reduces to the form of Eq.(20),
and the solution may take the form
\begin{equation}
F_{\pm kj}^>(r)=\sqrt{(E_k\pm\mu)k\over2E_k}
[\cos\eta_j(\pm E_k)J_m(kr)-\sin\eta_j(\pm E_k)N_m(kr)],
\end{equation}   
where the superscript ``$>$'' indicates $r>a$. Using Eqs.(16), (26),
and (29), one can obtain the expected asymptotic forms (30).
In the region $r<a$, Eq.(17) cannot be simplified. Let us
consider the behavior of the solution near $r=0$. We have
assumed that $V(r)$ is less singular than $r^{-1}$ when $r\to0$.
Thus, when $r\to0$, $V(r)$ is regular or behaves like $U_0/r^\delta$
where $U_0$ is a constant and $0<\delta<1$. Accordingly, Eq.(17)
becomes to leading terms
$$
F''+{1\over r}F'-{j_-^2\over r^2}F=0
\eqno(46{\rm a})
$$                        
in the first case or
$$
F''+{1+\delta\over r}F'-{j_-(j_-+\delta)\over r^2}F=0
\eqno(46{\rm b})
$$
in the second case. Therefore, the regular solution of Eq.(17) may
have the following power dependence on $r$ when $r\to0$:
$$
f_j^\pm (r,k)\to r^m
\eqno(47{\rm a})
$$
in the first case or
$$
f_j^\pm (r,k)\to \left\{
\begin{array} {l}
r^m,\quad j>0\\  \\
r^{m-\delta},\quad j<0
\end{array} 
\right.
\eqno(47{\rm b})
$$
in the second case, where we have denoted $F_{\pm kj}(r)$ with
these boundary conditions by $f_j^\pm (r,k)$, and $m=|j_-|$ as
before. The solution of Eq.(17) in the region $r<a$ is
\addtocounter{equation}{2}
\begin{equation}
F_{\pm kj}^<(r)=A_j^\pm (k)f_j^\pm(r,k),
\end{equation}
where the superscript ``$<$'' indicates $r<a$. In general the
coefficient $A_j^\pm$ depends on $k$ such that the two parts of
$F_{\pm kj}(r)$ can be appropriately connected at $r=a$. Obviously,
$F_{\pm kj}(r)$ and $G_{\pm kj}(r)$  should be continuous at $r=a$,
so that the probability density and the probability current density
are continuous at $r=a$. For simplicity, we assume that $V_a(r)$
is continuous at $r=a$, which means $V_a(r)\to0$ when $r\to a^-$
and $V_a(a)=0$. Then from Eq.(14) we see that $F_{\pm kj}'(r)$
is also continuous at $r=a$. Therefore, $F_{\pm kj}'(r)/F_{\pm kj}(r)$  
is continuous at $r=a$. This leads to
\begin{equation}
\tan\eta_j(\pm E_k)={\xi J'_m(\xi)-\beta_j^\pm(\xi)J_m(\xi)
\over \xi N'_m(\xi)-\beta_j^\pm(\xi)N_m(\xi)},
\end{equation}     
where $\xi=ka$ and
\begin{equation}
\beta_j^\pm(\xi)={af_j^{\pm\prime}(a,k)\over f_j^\pm(a,k)},
\end{equation}         
where the prime indicates differentiation with respect to $r$.
The above result shows that the behavior of $\eta_j(\pm E_k)$ is
determined by that of $\beta_j^\pm(\xi)$ and ultimately by that of
$f_j^\pm(a,k)$. The general dependence of $f_j^\pm(r,k)$ on $k$ may
be very complicated since Eq.(17) depends on $k$ in a rather
complicated way. This is quite different from the nonrelativistic case
where the function $f_m(r,k)$, the counterpart of $f_j^\pm(r,k)$, is
an integral function of $k$[16]. Fortunately, only the property of
$f_j^\pm(r,k)$ near $k=0$ is necessary for our purpose.

We consider the limit $E\to\mu$ of Eq.(17). In this limit it takes
the following form to the first order in $k^2$:
\begin{equation}
F''+P(r, k)F'+Q(r, k)F=0,
\end{equation}
where
$$
P(r,k)={1\over r}+{V'\over 2\mu-V}-{V'\over
2\mu(2\mu-V)^2}k^2,\eqno(51'{\rm a})$$
$$
Q(r,k)=-{j_-^2\over r^2}-2\mu V+V^2-{j_-V'\over
r(2\mu-V)}+\left[1-\frac V\mu+{j_-V'\over2\mu r(2\mu-V)^2}\right]k^2.
\eqno(51'{\rm b})
$$
We denote the solution of this equation that satisfies the boundary
condition (47) by $\tilde f_j^+(r,k)$. Note that both $P(r,k)$ and
$Q(r,k)$ are integral functions of $k$, and the boundary condition
(47) is independent of $k$. Then a theorem of Poincar\'e tells us
that $\tilde f_j^+(r,k)$ is also an integral function of $k$. On the
other hand, Eq.(17) coincides with Eq.(51) in the limit $E\to\mu$.
Therefore, $f_j^+(r,k)$ must coincide with $\tilde f_j^+(r,k)$ in the
limit $k\to0$ since they satisfy the same boundary condition. Hence
we conclude that $f_j^+(r,k)$ is an analytic function of $k$ in the
neighbourhood of $k=0$.
Moreover, $f_j^+(r,k)$ is an even functions of $k$, since Eq.(17)
is invariant under the change $k\to-k$, and the boundary condition is
independent of $k$. The above conclusion holds regardless of whether
$V(r)=V_a(r)$ or not.

We can now proceed as in the nonrelativistic case[16] and arrive
at the result
$$
\tan\eta_j(E_k)\to b_j^+ \xi^{2p_j^+}\quad{\rm or}\quad
{\pi\over 2\ln \xi}\quad(k\to 0), \eqno(52{\rm a})
$$
where $b_j^+\ne0$ is a contant and $p_j^+$ is a natural number.
For a strong repulsive potential Eq. (51) may become singular at the
points where $V=2\mu$. Then the above analysis is not reasonable.
In this case, however, we may consider $G_{kj}(r)$ instead of
$F_{kj}(r)$ and study the limit $E\to\mu$ of Eq. (19). The same result
(52a) can be attained in a similar way.
A similar analysis leads to
$$
\tan\eta_j(-E_k)\to b_j^- \xi^{2p_j^-}\quad{\rm or}\quad
{\pi\over 2\ln \xi}\quad(k\to 0),
\eqno(52{\rm b})
$$
where $b_j^-\ne0$ is a contant and $p_j^-$ is a natural number.
\addtocounter{equation}{1}

In the above analysis we have assumed that $V_a(r)$ is continuous at
$r=a$. This is, however, not mandatory. Using the condition that both
$F_{\pm kj}(r)$ and $G_{\pm kj}(r)$ are continuous at $r=a$, one can
arrive at the same result (52) even when $V(r)$ is not continuous
at $r=a$.

\section*{\centerline{\large IV. The Levinson theorem}}

With the above preparations,
we now proceed to establish the Levinson theorem by the Green
function method. We define the retarded Green function
$G({\bf r}, {\bf r'}, E)$ 
of the Dirac equation in an external potential $V({\bf r})$ by
\begin{equation}   
G({\bf r}, {\bf r'}, E)=\sum_\tau{\psi_\tau({\bf r})\psi^\dagger_\tau
({\bf r'})\over E-E_\tau+i\epsilon},
\end{equation}         
where $\{\psi_\tau({\bf r})\}$ is a complete set of orthonormal
solutions to Eq.(6) where $V$ may be a noncentral potential,
$E_\tau$ is the energy eigenvalue associated with the solution
$\psi_\tau({\bf r})$, and $\epsilon=0^+$.
$G({\bf r}, {\bf r'}, E)$ satisfies the equation
\begin{equation}
(E-H+i\epsilon)G({\bf r}, {\bf r'}, E)=\delta({\bf r}-{\bf r'}).
\end{equation}         
For a free particle, $V=0$. We denote the Hamiltonian by $H_0$ and
the solutions to the free Dirac equation by
$\psi_\tau^{(0)}({\bf r})$. The retarded Green function in this case
is defined as
\begin{equation}
G^{(0)}({\bf r}, {\bf r'}, E)=\sum_\tau{\psi^{(0)}_\tau({\bf r})
\psi^{(0)\dagger}_\tau
({\bf r'})\over E-E^{(0)}_\tau+i\epsilon},
\end{equation}         
where $E_\tau^{(0)}$  is the energy corresponding to the solution
$\psi^{(0)}_\tau({\bf r})$. 
$G^{(0)}({\bf r}, {\bf r'}, E)$ satisfies
\begin{equation}
(E-H_0+i\epsilon)G^{(0)}({\bf r}, {\bf r'}, E)
=\delta({\bf r}-{\bf r'}).
\end{equation}         
We have the integral equation for $G({\bf r}, {\bf r'}, E)$:
\begin{equation}
G({\bf r}, {\bf r'}, E)-G^{(0)}({\bf r}, {\bf r'}, E)=
\int d{\bf r}''\,G^{(0)}({\bf r}, {\bf r''}, E)V({\bf r''})
G({\bf r''}, {\bf r'}, E).
\end{equation}            

In a central field $V({\bf r})=V(r)$ [not necessarily $V_a(r)$],
we have discussed the solutions $\psi_\tau({\bf r})$ of the Dirac
equation (6) in Sec.II. They contain two different classes:
$\psi_{\pm kj}(r,\theta)$ with continuous energy spectrum $E_{\pm k}$
and $\psi_{\kappa j}(r,\theta)$ with discrete energy spectrum
$E_{\kappa j}$. However, we may discretize the continuous part of
the spectrum by requiring the radial wave function $F_{\pm kj}(r)$
or $v_{\pm kj1}(r)$ to vanish at a sufficiently large radius $R$
[$R\gg a$ for $V_a(r)$]. In this case we will denote all the
solutions by $\psi_{\kappa j}(r,\theta)$ and the corresponding
energies by $E_{\kappa j}$. We also apply the same prescription
to the free Dirac particle to turn the continuous energies
$E_{\pm k}^{(0)}$ into discrete ones $E_{\kappa j}^{(0)}$. The
waves functions will be denoted by $\psi_{\kappa j}^{(0)}(r,\theta)$
in the free case. We have then
\begin{equation}
G({\bf r}, {\bf r'}, E)=\sum_j{e^{ij_-(\theta-\theta')}\over2\pi}
\left(
\begin{array} {ll}
G_{j11}(r,r',E)&G_{j12}(r,r',E)e^{-i\theta'}\\ \\
G_{j21}(r,r',E)e^{i\theta}&G_{j22}(r,r',E)e^{i(\theta-\theta')}
\end{array} 
\right),
\end{equation}                    
where
$$
G_{jpq}(r,r',E)=\sum_{\kappa}{v_{\kappa jp}(r)v_{\kappa jq}^*(r')\over
\sqrt{rr'}(E-E_{\kappa j}+i\epsilon)},
\eqno(58')
$$                
where $p,q=1,2$ are used to denote spinor indices. For a free particle
we have similar relations. The integral equation for $G_{jpq}(r,r',E)$
may be derived from Eq.(57). The result turns out to be
\begin{equation}
G_{jpq}(r,r',E)-G_{jpq}^{(0)}(r,r',E)=\int dr''\,r''\sum_s
G_{jps}^{(0)}(r,r'',E)V(r'')G_{jsq}(r'',r',E).
\end{equation}                         

Since the wave function should be finite at $r=0$, we have the
boundary conditions
\begin{equation}
v_{\kappa jp}(0)=0, \quad v_{\kappa jp}^{(0)}(0)=0.
\end{equation}
For bound states with $|E_{\kappa j}|<\mu$, we have $v_{\kappa jp}
(\infty)=0$. For scattering states with $|E_{\kappa j}|>\mu$ we
may set $v_{\kappa jp}(r)=0$ when $r>R$.
However, it should be remarked
that in general $v_{\kappa j2}(R)\ne 0$ when we require $v_{\kappa
j1}(R)=0$. In any case, we have
$$
v_{\kappa jp}(\infty)=0, \quad v_{\kappa jp}^{(0)}(\infty)=0.
\eqno(60')
$$
Now the orthonormal relation takes the form ($27'$) for both
bound and scattering states, from which we have
\begin{equation}
(v_{\kappa' j},v_{\kappa j})=\delta_{\kappa \kappa'},\quad
\forall j,
\end{equation}                    
where
\begin{equation}
(v_{\kappa' j},v_{\kappa j})\equiv\int_0^\infty dr\,\sum_p
v_{\kappa' jp}^*(r)v_{\kappa jp}(r).
\end{equation}
As we have set $v_{\kappa jp}(r)=0$ when $r>R$ for scattering
states, the upper bound of integration in Eq.(62)  need not be
replaced by $R$ for these states. For the free particle we have
\begin{equation}
(v_{\kappa' j}^{(0)},v_{\kappa j}^{(0)})=\delta_{\kappa \kappa'},
\quad \forall j.
\end{equation}                      

Using Eqs.($58'$) and (61), it is easy to show that
\begin{equation}
\int dr\,r\sum_p G_{jpp}(r,r,E)=\sum_\kappa{1\over E-E_{\kappa j}
+i\epsilon}.
\end{equation}               
Employing the mathematical formula
\begin{equation}
{1\over x+i\epsilon}=P{1\over x}-i\pi\delta(x),
\end{equation}               
and taking the imaginary part of the above equation, we have
\begin{equation}
{\rm Im}\int dr\,r \sum_p G_{jpp}(r,r,E)=-\pi\sum_\kappa\delta
(E-E_{\kappa j}).
\end{equation}               
Integrating this equation over $E$ from $-\mu^-$ to $\mu^-$ yields
\begin{equation}
{\rm Im}\int_{-\mu^-}^{\mu^-}dE\,\int dr\,r\sum_p G_{jpp}(r,r,E)
=-n_j^-\pi,
\end{equation}               
where $n_j^-$ is the number of bound states in the angular momentum
channel $j$ with $|E_{\kappa j}|<\mu$, $\mu^-=\mu-0$, and $-\mu^-
=-\mu+0$. The existence of bound states with critical energies
$E_{\kappa j}=\pm\mu$ does not alter this result. Similarly, we can
show that
\begin{equation}
{\rm Im}\int_{-\mu^-}^{\mu^-}dE\,\int dr\,r\sum_p G_{jpp}^{(0)}
(r,r,E)=0.
\end{equation}               
Combining Eqs.(67) and (68) we obtain
\begin{equation}
{\rm Im}\int_{-\mu^-}^{\mu^-}dE\,\int dr\,r\sum_p[G_{jpp}(r,r,E)
-G_{jpp}^{(0)}(r,r,E)]=-n_j^-\pi.
\end{equation}               
On the other hand, substituting Eq.($58'$) and a similar one for
the free particle into the right-hand side (rhs) of Eq.(59) we
get
\begin{equation}
\int dr\,r\sum_p[G_{jpp}(r,r,E)-G_{jpp}^{(0)}(r,r,E)]
=\sum_{\kappa\kappa'}
{(v_{\kappa j}^{(0)}, Vv_{\kappa' j})(v_{\kappa' j},
v_{\kappa j}^{(0)})
\over (E-E_{\kappa j}^{(0)}+i\epsilon)(E-E_{\kappa' j}+i\epsilon)},
\end{equation}                      
where the definition of $(v_{\kappa' j},v_{\kappa j}^{(0)})$ is
similar to Eq.(62) and
\begin{equation}
(v_{\kappa j}^{(0)},Vv_{\kappa' j})\equiv\int_0^\infty dr\,\sum_p
v_{\kappa jp}^{(0)*}(r)V(r)v_{\kappa' jp}(r).
\end{equation}                                
Using the system of equations (15) and a similar one for the free
case, and employing the boundary conditions (60) and ($60'$), one
can show that
\begin{equation}
(v_{\kappa j}^{(0)},Vv_{\kappa' j})=(E_{\kappa' j}-E_{\kappa j}^{(0)})
(v_{\kappa j}^{(0)},v_{\kappa' j}).
\end{equation}                      
Substituting this result into Eq.(70) and taking the imaginary part,
we get
\begin{equation}
{\rm Im}\int dr\,r\sum_p[G_{jpp}(r,r,E)-G_{jpp}^{(0)}(r,r,E)]
=\pi\sum_{\kappa\kappa'}
[\delta(E-E_{\kappa j}^{(0)})-\delta(E-E_{\kappa' j})]
|(v_{\kappa j}^{(0)}, v_{\kappa' j})|^2.
\end{equation}                      
Integrating this equation over $E$ from $-\infty$ to $+\infty$ it
turns out that
\begin{equation}
{\rm Im}\int_{-\infty}^{+\infty}dE\,\int dr\,r
\sum_p[G_{jpp}(r,r,E)-G_{jpp}^{(0)}(r,r,E)]
=0.
\end{equation}                      
As in the nonrelativistic case, this means that the total number of
states in a specific angular momentum channel is not altered by
an external field, except that some scattering states are ``pulled
down'' into the bound-state region. Here the external field may be
either attractive or repulsive. This is different from the
nonrelativistic case, where bound states exist only in attractive
fields. Another new character of Dirac particles is that when the
potential, say, an attractive potential, becomes strong enough, some
bound states may even be pulled down into the region of
negative-energy scattering states. This new character, however, does
not alter the above conclusion that the total number of states in a
specific angular momentum channel remains unchanged, since the
conclusion is a consequence of the Dirac equation (15) and the
completeness of the whole set of states.
If the integration over $E$ in Eq.(74) is performed from
$-\mu^-$ to $\mu^-$, we have
\begin{equation}
{\rm Im}\int_{-\mu^-}^{\mu^-}dE\,\int dr\,r
\sum_p[G_{jpp}(r,r,E)-G_{jpp}^{(0)}(r,r,E)]
=-\pi\sum_{\kappa'}{}'\sum_\kappa
|(v_{\kappa' j}, v_{\kappa j}^{(0)})|^2,
\end{equation}                      
where we have taken into account the fact that $|E_{\kappa j}^{(0)}|
>\mu$, and the prime in $\sum'_{\kappa'}$ indicates
that the summation over $\kappa'$ is performed only for bound states
with $|E_{\kappa' j}|<\mu$. Using the completeness relation (28),
which now takes the form
$$
\sum_{\kappa j}\psi_{\kappa j}({\bf r})
\psi_{\kappa j}^\dagger({\bf r'})
=\delta ({\bf r}-{\bf r'}),
\eqno(28')
$$
where the summation over $\kappa$ is performed for all states, we
can show that
\begin{equation}
\sum_\kappa v_{\kappa jp}^{(0)}(r)v_{\kappa jq}^{(0)*}(r')
=\delta_{pq}\delta(r-r'),\quad \forall j.
\end{equation}                    
This in turn leads to
\begin{equation}
\sum_\kappa |(v_{\kappa' j}, v_{\kappa j}^{(0)})|^2=1,\quad
\forall \kappa',j.
\end{equation}                    
Substituting Eq.(77) into Eq.(75) we recover Eq.(69). Combining
Eqs.(69) and (74) we obtain
\begin{equation}
{\rm Im}\left[\int_{-\infty}^{-\mu^-}+\int_{\mu^-}^{+\infty}\right]
dE\,\int dr\,r\sum_p[G_{jpp}(r,r,E)
-G_{jpp}^{(0)}(r,r,E)]=n_j^-\pi.
\end{equation}               
We have thereupon finished the first step in our establishment of
the Levinson theorem.

The next step  is to calculate the lhs of Eq.(78) in another way. In
the above treatment we have discretized the continuous spectrum of
$E_{\kappa j}^{(0)}$ and the continuous part of $E_{\kappa j}$. In the
following we will directly deal with these continuous spectra, and
use the notations of Sec.II. Then the retarded Green function
$G({\bf r,r'}, E)$ is given by Eq.(58) but where $G_{jpq}(r,r',E)$
is given by
\begin{eqnarray}
G_{jpq}(r,r',E)&=& 
\int_0^\infty dk\,\left[{v_{kjp}(r)v_{kjq}^*(r')\over
\sqrt{rr'}(E-E_{kj}+i\epsilon)}+
{v_{-kjp}(r)v_{-kjq}^*(r')\over
\sqrt{rr'}(E-E_{-kj}+i\epsilon)}\right]\nonumber \\
&+&\sum_{\kappa}{v_{\kappa jp}(r)v_{\kappa jq}^*(r')\over
\sqrt{rr'}(E-E_{\kappa j}+i\epsilon)},
\end{eqnarray}              
where the integration is performed over scattering states while
the summation over bound states, and $E_{\pm kj}=E_{\pm k}$ [cf.
Eq.(21)] is independent of $j$. For $G_{jpq}^{(0)}(r,r',E)$
we have a similar expression but without the last term. Using the
formula (65) it is easy to show that
\begin{eqnarray}
\lefteqn{}&&{\rm Im}\int dr\,r\sum_p G_{jpp}(r,r,E)\nonumber \\
\lefteqn{}&&=-\pi\int_0^\infty dk\,[\delta(E-E_{kj})(v_{kj}, v_{kj})
+\delta(E-E_{-kj})(v_{-kj}, v_{-kj})] 
-\pi\sum_\kappa\delta(E-E_{\kappa j}).~~~~~~~~
\end{eqnarray}             
Here the inner products $(v_{kj},v_{kj})$ etc. are defined in the
same way as Eq.(62).
Integrating this equation over $E$ as follows we have
\begin{eqnarray}
&&{\rm Im}\left[\int_{-\infty}^{-\mu^-}
+\int_{\mu^-}^{+\infty}\right]dE\,\int dr\,r\sum_p G_{jpp}(r,r,E)
\nonumber \\
&&=-\pi\int_0^\infty dk\,[(v_{kj}, v_{kj})
+(v_{-kj}, v_{-kj})]
-\pi(\delta_{\mu j}+\delta_{-\mu j}),
\end{eqnarray}             
where $\delta_{\mu j}=1$ ($\delta_{-\mu j}=1$) if there exists a
bound state in the angular momentum  channel $j$ with critical
energy $E_{\kappa j}=\mu$ ($E_{\kappa j}=-\mu$), otherwise 
$\delta_{\mu j}=0$ ($\delta_{-\mu j}=0$). (See Sec.V for
further discussions.) Similar to Eq.(81) we have for the free case
\begin{equation}
{\rm Im}\left[\int_{-\infty}^{-\mu^-}+\int_{\mu^-}^{+\infty}\right]
dE\,\int dr\,r\sum_p G_{jpp}^{(0)}(r,r,E)=
-\pi\int_0^\infty dk\,[(v_{kj}^{(0)}, v_{kj}^{(0)})
+(v_{-kj}^{(0)}, v_{-kj}^{(0)})].
\end{equation}             
Combining Eqs.(81) and (82) we obtain
\begin{eqnarray}
&&{\rm Im}\left[\int_{-\infty}^{-\mu^-}
+\int_{\mu^-}^{+\infty}\right]
dE\,\int dr\,r\sum_p[G_{jpp}(r,r,E)
-G_{jpp}^{(0)}(r,r,E)]\nonumber \\
&&=-\pi\int_0^\infty dk\,[(v_{kj}, v_{kj})
-(v_{kj}^{(0)}, v_{kj}^{(0)})]
-\pi\int_0^\infty dk\,[(v_{-kj}, v_{-kj})-(v_{-kj}^{(0)},
v_{-kj}^{(0)})]\nonumber \\
&&-\pi(\delta_{\mu j}+\delta_{-\mu j}).
\end{eqnarray}             
From the orthonormal relations (23) and (27) one can show that
\begin{equation}
(v_{\pm k'j}, v_{\pm kj})=\delta(k-k'),\quad
(v_{\pm k'j}^{(0)}, v_{\pm kj}^{(0)})=\delta(k-k'),
\quad \forall j.
\end{equation}   
This implies that both integrands in the rhs of Eq.(83) are
$\delta(0)-\delta(0)$. There is, however, a subtle difference between
these two $\delta(0)$'s, and it is this difference that leads to the
Levinson theorem. To get rid of the difficulty of infiniteness,
we define
\begin{equation}
(v_{lj},v_{kj})_{r_0}\equiv\int_0^{r_0} dr\,\sum_p
v_{ljp}^*(r)v_{kjp}(r),
\end{equation}                    
and obtain $(v_{kj},v_{kj})$ in the limit $l\to k$ and $r_0\to\infty$.
Using two systems of equations satisfied by $v_{kjp}$ and $v_{ljp}$
[cf. Eq.(15)], and the boundary conditions [cf. Eq.(60)]
\begin{equation}
v_{kjp}(0)=0,\quad v_{kjp}^{(0)}(0)=0,
\end{equation}                 
it can be shown that
\begin{equation}
(E_{lj}-E_{kj})(v_{lj},v_{kj})_{r_0}=v_{lj2}^*(r_0)v_{kj1}(r_0)
-v_{lj1}^*(r_0)v_{kj2}(r_0).
\end{equation}
Since $r_0$ is large, we can use Eq.(30) to evaluate the rhs and
in the limit $l\to k$ we get
\begin{equation}
(v_{kj}, v_{kj})_{r_0}={r_0\over \pi}+{1 \over \pi}{d\eta_j(E_k)
\over dk}-
(-)^m{\mu\over 2\pi kE_k}\cos[2kr_0+2\eta_j(E_k)].
\end{equation}      
Similarly, we have
\begin{equation}
(v_{kj}^{(0)}, v_{kj}^{(0)})_{r_0}={r_0\over \pi}-
(-)^m{\mu\over 2\pi kE_k}\cos 2kr_0.
\end{equation}      
Obviously, the infiniteness lies in the first term $r_0/\pi$ when
we take the limit $r_0\to\infty$. This disappears when we subtract
Eq.(89) from Eq.(88). 
Using the well-known formulas
\begin{equation}
\lim_{r_0\to\infty}{\sin 2kr_0\over \pi k}=\delta(k),
\end{equation}   
and $g(k)\delta(k)=g(0)\delta(k)$ for any continuous function $g(k)$,
we obtain
\begin{eqnarray}
&&(v_{kj}, v_{kj})_{r_0}-(v_{kj}^{(0)}, v_{kj}^{(0)})_{r_0}
\nonumber \\
&&={1\over\pi}{d\eta_j(E_k)\over dk}+{(-)^m\over 2}\delta(k)
\sin2\eta_j(\mu)+
(-)^m{\mu\over\pi kE_k}\cos 2kr_0\sin^2\eta_j(E_k).
\end{eqnarray}    
In a similar way, we can show that
\begin{eqnarray}
&&(v_{-kj}, v_{-kj})_{r_0}-(v_{-kj}^{(0)}, v_{-kj}^{(0)})_{r_0}
\nonumber \\
&&={1\over\pi}{d\eta_j(-E_k)\over dk}-{(-)^m\over 2}\delta(k)
\sin2\eta_j(-\mu)-
(-)^m{\mu\over\pi kE_k}\cos 2kr_0\sin^2\eta_j(-E_k).
\end{eqnarray}    
So far in this section $V(r)$ need not be $V_a(r)$. In the following
we set $V(r)=V_a(r)$. Then Eq.(52) holds. Therefore
$\sin2\eta_j(\mu)=\sin2\eta_j(-\mu)=0$.
Integrating Eq.(91) over $k$
(from 0 to $+\infty$) and taking the limit $r_0\to \infty$ we have
\begin{eqnarray}
&&\int_0^\infty dk\,[(v_{kj}, v_{kj})-(v_{kj}^{(0)},v_{kj}^{(0)})]
\nonumber \\
&&={1\over\pi}[\eta_j(+\infty)-\eta_j(\mu)]+
(-)^m{\mu\over\pi}\lim_{r_0\to\infty}\int_0^\infty dk\,
{\sin^2\eta_j(E_k)\over kE_k}\cos 2kr_0.
\end{eqnarray}    
The last term in this equation can be decomposed into two integrals,
the first from 0 to $\varepsilon=0^+$, while the second
from $\varepsilon$ to $+\infty$.
The second integral vanishes in the limit $r_0\to\infty$
since the factor $\cos2kr_0$ oscillates very rapidly.
For the first integral, we have
$$ \int_0^\varepsilon dk\,
{\sin^2\eta_j(E_k)\over kE_k}\cos 2kr_0={1\over \mu}
\int_0^\varepsilon dk\,
{\sin^2\eta_j(E_k)\over k}={1\over \mu}
\int_0^{\varepsilon a} d\xi\,{\sin^2\eta_j(E_k)\over \xi}$$
as $k$ is very small. For the same reason we can use Eq. (52a). In
the first case, $\tan\eta_j(E_k)=b_j^+\xi^{2p_j^+}$, we have
$\sin^2\eta_j(E_k)=b_j^{+2}\xi^{4p_j^+}$, so that
$${1\over\mu}
\int_0^{\varepsilon a} d\xi\,{\sin^2\eta_j(E_k)\over \xi}
={b_j^{+2}\over\mu}
\int_0^{\varepsilon a} d\xi\,\xi^{4p_j^+ -1}=
{b_j^{+2}\over 4\mu p_j^+}(\varepsilon a)^{4p_j^+}\to 0,
\quad (\varepsilon\to 0^+).$$
In the second case, $\tan\eta_j(E_k)=\pi/2\ln\xi$, we have
$\sin^2\eta_j(E_k)=\pi^2/4\ln^2\xi$, so that
$${1\over\mu}
\int_0^{\varepsilon a} d\xi\,{\sin^2\eta_j(E_k)\over \xi}
={\pi^2\over 4\mu}
\int_0^{\varepsilon a} d\xi\,{1\over \xi\ln^2\xi}=-
{\pi^2\over 4\mu}{1\over\ln\varepsilon a}\to 0,
\quad (\varepsilon\to 0^+),$$
where the final integral is calculated by setting $\ln\xi=w$.
Thus the first integral vanishes as well.
Therefore, we have
\begin{equation}
\int_0^\infty dk\,[(v_{kj}, v_{kj})-(v_{kj}^{(0)},v_{kj}^{(0)})]
={1\over\pi}[\eta_j(+\infty)-\eta_j(\mu)].
\end{equation}    
On the basis of Eqs.(92) and (52b) we can show that
\begin{equation}
\int_0^\infty dk\,[(v_{-kj}, v_{-kj})-(v_{-kj}^{(0)},v_{-kj}^{(0)})]
={1\over\pi}[\eta_j(-\infty)-\eta_j(-\mu)].
\end{equation}    
Substituting Eqs.(94) and (95) into Eq.(83) we obtain
\begin{eqnarray}
&&{\rm Im}\left[\int_{-\infty}^{-\mu^-}
+\int_{\mu^-}^{+\infty}\right]
dE\,\int dr\,r\sum_p[G_{jpp}(r,r,E)-G_{jpp}^{(0)}(r,r,E)]\nonumber \\
&&=[\eta_j(\mu)-\eta_j(+\infty)]+[\eta_j(-\mu)-\eta_j(-\infty)]
-\pi(\delta_{\mu j}+\delta_{-\mu j}).
\end{eqnarray}             
Combining this result with Eq.(78) we arrive at
\begin{equation}
[\eta_j(\mu)-\eta_j(+\infty)]+[\eta_j(-\mu)-\eta_j(-\infty)]
=n_j\pi,
\end{equation}      
where
\begin{equation}
n_j=n_j^-+\delta_{\mu j}+\delta_{-\mu j}
\end{equation}
is the total number of bound states, including the possible ones
with critical energies $\pm\mu$, in the angular momentum channel $j$.
Eq.(97) is the Levinson theorem for Dirac particles in an external
central field in two dimensions.
In the next section we will discuss some relevant points of the
theorem.

\section*{\centerline{\large V. Discussions}}

{\it 1. On bound states with critical energies}.
We consider the cut-off potential $V_a(r)$. In the region $r>a$,
$V_a(r)=0$. Equations (16) and (17) are applicable for $E=\mu$,
while for $E=-\mu$ Eqs.(18) and (19) should be employed.

In the region $r>a$ and for $E=\mu$, Eq.(17) becomes
\begin{equation}
F''+{1\over r}F'-{m^2\over r^2}F=0,
\end{equation}
where $m=|j_-|$ as before. The well behaved solution is
\begin{equation}
F_{\mu j}^>(r)=r^{-m}.
\end{equation}      
The corresponding solution $G_{\mu j}^>(r)$ obtained from Eq.(16) 
decreases more rapidly than
$F_{\mu j}^>(r)$ when $r\to\infty$. Thus the normalizability of the
solution is determined by the behavior of $F_{\mu j}^>(r)$ at large
$r$. Obviously, the solution can be normalized only when $m>1$.
It is well behaved when $m=0,1$, but cannot be normalized. In other
words, the solution with energy $E=\mu$ is a bound state only when
$j>3/2$ or $j<-1/2$. The cases $j=\pm1/2, 3/2$ do not correspond to
bound states. Of course, a bound state with $E=\mu$ actually exists
only when the potential $V_a(r)$ has a specific form such that the
solution is continuous at $r=a$. For a square-well potential with
depth $V_0$, this leads to
$$
k_0aJ_{m+1}(k_0a)=\left(2m+{mV_0\over\mu}\right)J_m(k_0a)
\eqno(101{\rm a})
$$
for $j>3/2$ or
$$ J_{m-1}(k_0a)=0 \eqno(101{\rm b}) $$
for $j<-1/2$, where $k_0=\sqrt{V_0^2+2\mu V_0}$. For a given $j$, this
is satisfied only for some specified depth $V_0$ or radius $a$.
\addtocounter{equation}{1}

For $E=-\mu$, we consider Eq.(19). In the region $r>a$ it reduces
to
\begin{equation}
G''+{1\over r}G'-{m'^2\over r^2}G=0,
\end{equation}                   
where $m'=|j_+|$. The well behaved solution is
\begin{equation}
G_{-\mu j}^>(r)=r^{-m'}.
\end{equation}      
One can get $F_{-\mu j}^>(r)$ from Eq.(18). It decreases
more rapidly than $G_{-\mu j}^>(r)$ when $r\to\infty$. Therefore the
solution is normalizable when $m'>1$. In other words, the solution is
a bound state when $j>1/2$ or $j<-3/2$. The cases $j=\pm1/2,-3/2$ do
not correspond to bound states. For the square-well potential, a bound
state with $E=-\mu$ really exists only when $V_0>2\mu$ and satisfies
$$
\tilde k_0aJ_{m'+1}(\tilde k_0a)=\left(2m'-{m'V_0\over\mu}
\right)J_{m'}(\tilde k_0a)
\eqno(104{\rm a})
$$
for $j<-3/2$ or
$$ J_{m'-1}(\tilde k_0a)=0 \eqno(104{\rm b}) $$
for $j>1/2$, where $\tilde k_0=\sqrt{V_0^2-2\mu V_0}$.
Given $j$, say $j=-5/2$, there exist infinitely many solutions of
$V_0$ to Eq. (101b). They are functions of $a$. In general most of
them cannot satisfy Eq. (104a) at the same time as the two equations
are independent, thus the two critical energy bound states do not
appear simultaneously. However, the solutions of Eq. (104a) are also
functions of $a$. By varying the parameter $a$, it may be possible to
match some specific solution of Eq. (101b) with some specific one of
Eq. (104a). When this really happens, the two critical energy bound
states can appear simultaneously. Anyway, it cannot be asserted that
the two critical energy bound states never appear simultaneously
for any potential.

\addtocounter{equation}{1}

The form of the Levinson theorem (97) is not modified by the existence
of the critical energy states, regardless of whether they are bound
states or not. The existence of critical energy bound states just
changes $n_j$ from $n_j^-$ to $n_j^-+1$ (when there is one with
$E=\mu$ or $E=-\mu$) or $n_j^-+2$ (when the two appear
simultaneously), and does not alter the form of Eq. (97). In three
dimensions, the theorem involves additional terms that vanish except
when there exist half bound states[10]. In two dimensions there is no
such term, which is clear from the result (97). Thus the theorem in
two dimensions is not affected by the existence of half bound states.
The reason for the difference between two- and
three-dimensional cases is similar to that in the nonrelativistic
theory. This has been discussed in detail in Ref. [16].
                                    
{\it 2. About $\eta_j(\pm\infty)$}.  We write down two systems of
equations for the radial wave functions $v_{kjp}^U(r)$ and
$\tilde v_{kjp}^U(r)$ in two external fields
$U(r)$ and $\tilde U(r)$, respectively. Using the boundary condition
(86) and the asymptotic form (30), it is not difficult to show that
\begin{equation}
\sin[\tilde\eta_j^U(E_k)- \eta_j^U(E_k)]=-{\pi E_k\over k}
(\tilde v_{kj}^U, (\tilde U-U) v_{kj}^U).
\end{equation}    
Here the superscript $U$ is used to distinguish the wave functions
and phase shifts from those in the external field $V(r)$. 
Now we set $U(r)=\lambda V(r)$ [$V(r)$ not necessarily be $V_a(r)$],
$\tilde U(r)=(\lambda+\Delta\lambda)V(r)$,
and denote $v_{kjp}^U(r)=v_{kjp}(r,\lambda)$,
$\eta_j^U(E_k)=\eta_j(E_k,\lambda)$. When $\Delta\lambda\to0$,
$\tilde v_{kjp}^U(r)=v_{kjp}(r,\lambda+\Delta\lambda)$ can be
replaced by $v_{kjp}(r,\lambda)$, and Eq.(105) becomes
\begin{equation}
\sin[\Delta\eta_j(E_k,\lambda)]=-{\pi E_k\over k}\Delta\lambda
(v_{kj}(r,\lambda),V(r)v_{kj}(r,\lambda)),
\end{equation}    
where $\Delta\eta_j(E_k,\lambda)=\eta_j(E_k,\lambda+\Delta\lambda)-
\eta_j(E_k,\lambda)$. Obviously, $v_{kjp}(r,0)=v_{kjp}^{(0)}(r)$,
$v_{kjp}(r,1)=v_{kjp}(r)$, $\eta_j(E_k,1)=\eta_j(E_k)$. It is natural
to define $\eta_j(E_k,0)=0$ in the absence of an external field.
It is also natural to require that for any finite $k$,
$\eta_j(E_k,\lambda)$ be continuous functions
of $\lambda$ when $\lambda$ varies continuously from 0 to 1,
as all quantities in the equation and boundary conditions are
continuous in $\lambda$.
With this requirement and the above definition $\eta_j(E_k,0)=0$
one can determine $\eta_j(E_k)$. It should be remarked, however,
that $\eta_j(\mu,\lambda)\equiv\lim_{k\to 0}\eta_j(E_k,\lambda)$
is not continuous in $\lambda$. Otherwise the Levinson
theorem would be impossible. To see this, let us have a look at the
original Levinson theorem in an attractive potential $\lambda V(r)$
in three dimensions:
\begin{equation}
\delta_l(0,\lambda)-\delta_l(\infty,\lambda)=n_l(\lambda)\pi,
\quad l=0,1,2,\ldots,
\end{equation}     
where we have not included the modified $l=0$ case. If
$\delta_l(0,\lambda)$ is a continuous function of $\lambda$, then
the lhs of Eq.(107) is continuous in $\lambda$. On the other hand,
$n_l(\lambda)$ is a nonnegative integer. When $\lambda$ varies
continuously from 0 to 1, $n_l(\lambda)$ varies from 0 to $n_l$ by
discontinuous jumps. The rhs of Eq.(107) is obviously not a continuous
function of $\lambda$. This is a contradiction. So that
$\delta_l(0,\lambda)$ cannot be continuous in $\lambda$.
The case for $\eta_j(\mu, \lambda)$ is similar.

Since $\eta_j(E_k,\lambda)$ is a continuous function of $\lambda$,
the $\sin[\Delta\eta_j(E_k,\lambda)]$ on the lhs of Eq.(106) can be
safely replaced by $\Delta\eta_j(E_k,\lambda)$, and we have
\begin{equation}
{d\eta_j(E_k,\lambda)\over d\lambda}=-{\pi E_k\over k}
(v_{kj}(r,\lambda),V(r)v_{kj}(r,\lambda)).
\end{equation}    
When $E_k$ is infinitely large, since $V(r)$ is not very singular
at $r=0$ and is regular elsewhere, we may ignore $\lambda V(r)$
in the system of radial equations and approximately replace
$v_{kj}(r,\lambda)$ by $v_{kj}^{(0)}(r)$ in the above equation.
Then Eq.(108) can be easily integrated over $\lambda$ from 0
to 1 and results in
\begin{equation}
\eta_j(E_k)=-{\pi E_k\over k}
(v_{kj}^{(0)},Vv_{kj}^{(0)}).
\end{equation}    
Substituting the exact solution (22) into Eq.(109) and then
replacing the Bessel functions by their asymptotic forms as $k$ is
large,  we arrive at
$$
\eta_j(+\infty)=-\int_0^\infty dr\,V(r).\eqno(110{\rm a})
$$
In a similar way we can show that
$$
\eta_j(-\infty)=\int_0^\infty dr\,V(r).\eqno(110{\rm b})
$$
As we have assumed that $V(r)$ is less singular than $r^{-1}$ when
$r\to 0$ and decreases more rapidly than $r^{-2}$ when $r\to\infty$,
the above integrals converge. These
results have the same form as those in three dimensions[9,17,18].
Of course, Eq.(110) holds in the special case $V(r)=V_a(r)$. 
As a consequence of the above results, we have
\addtocounter{equation}{1}
\begin{equation}
\eta_j(+\infty)+\eta_j(-\infty)=0.
\end{equation}     
This reduces the Levinson theorem (97) to the form of Eq.(1).
It means that the sum of the phase shifts at the two thresholds
serves as a counter for the bound states in a specific angular
momentum channel. This is similar to the case in three dimensions, but
is somewhat simpler.

The Levinson theorem for Dirac particles could not be separated into
two parts, each of which likes that for Schr\"odinger particles
[6,9,10,19]. Essentially this is because that positive-energy
solutions or negative-energy solutions alone do not form a complete
set. An evidence can be seen as follows. In an attractive potential,
say, Eq. (108) shows that $\eta_j(E_k,\lambda)$ is positive and
increases with $\lambda$. Similarly, it can be shown that
$\eta_j(-E_k,\lambda)$ is negative and decreases when $\lambda$
increases. Thus $\eta_j(-\mu)$ may be negative when the attractive
potential becomes strong enough. In three dimensions this has been
verified by numerical calculations[10,18]. As $\eta_j(-\mu)$ may be
negative, it cannot always equal a nonnegative integer (in unit of
$\pi$) and thus cannot serve as a counter.

{\it 3. Extension to more general potentials}. Throughout this paper
we have assumed that $V(r)$ is less singular than $r^{-1}$ when
$r\to 0$ and decreases more rapidly than $r^{-2}$ when $r\to\infty$.
In the development of the Levinson theorem, we further cut off
$V(r)$ to the special case $V_a(r)$ for the sake of exact analysis.
However, the radius
$a$ beyond which $V_a(r)$ vanishes is not specified in our discussion.
Though both sides of Eq.(1) depend on the particular
form of $V_a(r)$ and thus depend on $a$, the equality between them
does
not. Thus we expect that Eq.(1) remains valid when $V_a(r)$ is varied
continuously to the limit $a\to\infty$ provided that the asymptotic
form (30) holds and $n_j$ remains finite in the
process. This requires that $V(r)$ decreases rapidly enough when
$r\to\infty$. It seems that the Levinson theorem holds at least
for short-range potentials that decrease more rapidly than
$r^{-2}$ when $r\to \infty$. 

\section*{\centerline{\large Acknowledgment}}

The author is grateful to Professor Guang-jiong Ni for useful
communications, for discussions, and for continuous encouragement.
This work was supported by the
National Natural Science Foundation of China.


\newpage
\begin{thebibliography}{99}
\bibitem{1}N. Levinson, K. Dan. Vidensk. Selsk. Mat.-Fys. Medd. 25,
           No.9 (1949).

\bibitem{2}R. G. Newton, J. Math. Phys. 1, 319 (1960).

\bibitem{3}J. M. Jauch, Helv. Phys. Acta. 30, 143 (1957).

\bibitem{4} A. Martin, Nuovo Cimento 7, 607 (1958).

\bibitem{5} R. G. Newton, J. Math. Phys. 18, 1348; 1582 (1977);
            {\it Scattering Theory of Waves and Particles}
            (McGraw-Hill, New York, 1966).

\bibitem{6} G.-J. Ni, Phys. Energ. Fort. Phys. Nucl. 3, 432 (1979).

\bibitem{7}Z.-Q. Ma, J. Math. Phys. 26, 1995 (1985); Phys. Rev. D 33,
           1745 (1986).

\bibitem{8}Z. R. Iwinski, L. Rosenberg, and L, Spruch, Phys. Rev.
           A 31, 1229 (1985); Phys. Rev. Lett. 54, 1602 (1985).

\bibitem{9}M. C. Barth\'el\'emy, Ann. Inst. Henri Poincar\'e 7,
           115 (1967).

\bibitem{10}Z.-Q. Ma and G.-J. Ni, Phys. Rev. D 31, 1482 (1985).

\bibitem{11}Z.-Q. Ma, Phys. Rev. D 32, 2203; 2213 (1985).

\bibitem{12}Y.-G. Liang and Z.-Q. Ma, Phys. Rev. D 34, 565 (1986).

\bibitem{13} N. Poliatzky, Phys. Rev. Lett. 70, 2507 (1993);
            Helv. Phys. Acta. 66, 241 (1993).
\bibitem{14} J. Piekarewicz, Phys. Rev. C 48, 2174 (1993).
\bibitem{15} R. Jackiw and G. Woo, Phys. Rev. D 12, 1643 (1975);
B. Berg, M. Karowski, W. R. Theis, and H. J. Thun, Phys. Rev. D 17,
1172 (1978); A. J. Niemi and G. W. Semenoff, Phys. Rev. D 32, 471
(1985).
\bibitem{16}Q.-G. Lin, Phys. Rev. A 56, 1938 (1997).
\bibitem{17}F. Calogero, {\it Variable Phase Approach to Potential
Scattering} (Academic, New York, 1967).
\bibitem{18}G.-J. Ni and S.-Q. Chen, {\it Levinson Theorem, Anomaly,
and the Phase Transition of Vacuum} (Shanghai Scientific \& Technical,
Shanghai, 1995).
\bibitem{19}Z.-Q. Ma, Phys. Rev. Lett. 76, 3654 (1996).
\end{thebibliography}
\end{document}